\documentclass{emulateapj}
\begin{document}

\def\arcdeg{\hbox{$^\circ$}}
\def\arcmin{\hbox{$^\prime$}}
\def\arcsec{\hbox{$^{\prime\prime}$}}
\def\lesssim{\mathrel{\hbox{\rlap{\hbox{\lower4pt\hbox{$\sim$}}}\hbox{$<$}}}}
\def\gtrsim{\mathrel{\hbox{\rlap{\hbox{\lower4pt\hbox{$\sim$}}}\hbox{$>$}}}}

\title{Submillimeter imaging of RCSJ022434-0002.5: Intense activity in a high redshift cluster? }

\shorttitle{Submillimeter imaging of RCSJ022434-002.5}
\shortauthors{Webb et al.}

\author{T.M.A. Webb\altaffilmark{1}, H.K.C. Yee\altaffilmark{2}, R.J. Ivison\altaffilmark{3}, H. Hoekstra\altaffilmark{4}, M.D. Gladders\altaffilmark{5},  L.F. Barrientos\altaffilmark{6}, Hsieh, B.C.\altaffilmark{7} }

\altaffiltext{1}{Leiden Observatory, University of Leiden, Niels Bohrweg 2, 2333 CA Leiden, The Netherlands}

\altaffiltext{2}{Department of Astronomy and Astrophysics, University of Toronto, 60 St George St, Toronto, 
Ontario, Canada, M5S 1A1}

\altaffiltext{3}{Astronomy Technology Centre, Royal Observatory, Blackford Hill, Edinburgh, UK, EH9 3HJ}

\altaffiltext{4}{Department of Physics \& Astronomy, University of Victoria, Elliott Building, 3800 Finnerty Rd., Victoria, BC, V8P 5C2 Canada}

\altaffiltext{5}{Carnegie Observatories, 813 Santa Barbara Street, Pasadena, California, 91101, USA} 

\altaffiltext{6}{Departamento de Astronomia y Astrofisica, Facultad de Fisica, Pontificia Universidad Catolica de Chile, Casilla 306, Santiago, Chile}

\altaffiltext{7}{ Institute of Astronomy, National Central University, No. 300, Jhongda Rd, 
Jhongli City, Taoyuan County 320, Taiwan, R.O.C.;
Institute of Astronomy and Astrophysics, Academia Sinica.
P.O. Box 23-141, Taipei 106, Taiwan, R.O.C.}

\begin{abstract}
We present deep 850{\micron} imaging of the $z$=0.773 strong lensing galaxy cluster  RCSJ022434-0002.5 from the Red-Sequence Cluster Survey (RCS). These data are part of a larger submillimeter survey of RCS clusters, with SCUBA on the JCMT. We find five objects at 850{\micron}, all of which are also detected at either 1.4-GHz, 450$\mu$m or both. The number density of objects in this field is in general agreement with the blank-field source counts; however, when combined with other cluster surveys a general tendency of cluster fields towards higher submm number densities is seen, which may be the result of unrecognized submillimeter luminous cluster galaxies. Primarily employing optical photometric redshifts  we show that two of the five submillimeter galaxies in this field are consistent with being cluster members, while two are more likely background systems.

\end{abstract}


\section{Introduction}

The study of galaxy clusters encompasses  almost every aspect of cosmology, from the determination of fundamental constants to the influence of the local environment on galaxy evolution. Yet a solid understanding of the process of cluster evolution has remained elusive. 
In a hierarchical universe clusters begin to form at very high redshift, then grow and evolve through cluster merging and accretion of groups.
Understanding this process,which may continue to significantly lower redshifts than the original epoch of formation, and its importance in driving cluster galaxy evolution,  will be a major step towards clarifying the complex physical details of hierarchical structure formation, at the level of individual objects.

By a redshift of $z\sim$ 1 galaxy clusters show already significant differences  from their local descendants, indicative of younger and more active systems \citep[e.g.,][]{but84,van99,van01,dwa99,sma99,bes02a}.  As in the field, much of this activity is dust enshrouded \citep{pog99,sma99} and conclusions drawn solely from optical studies are potentially misleading or at best incomplete. Indeed, recent observations of moderate redshift clusters ($z\sim$ 0.2-0.4) with ISO at 15$\mu$m, which directly samples hot dust,  have detected  dusty starburts  \citep{fad00,duc02,coi04}, and it is estimated for one cluster that perhaps 90\% of the star formation is dust enshrouded and undetected in  optical studies \citep{duc02}.
The ISO results are particularly interesting since, at least for one cluster, A1689 \citep{fad00}, the fraction of dusty galaxies is greater than in the field,  implying  the cluster environment is somehow responsible for triggering the intense  star formation.

Thus far,  extreme star-formation activity in clusters (such as that  seen in ultraluminous infrared galaxies [ULIRGs] at high and low redshift) has not been unambiguously detected; even the systems seen  by ISO  are limited to rates of a few tens of solar masses per year (assuming no contribution from active galactic  nuclei [AGN]). The exception to this is the handful of central cD galaxies detected at 850$\mu$m primarily during surveys for faint background sources \citep[K.K. Knudsen et al.~in preparation]{edg99,sma02,cow02}.  \citet{bes02b} reported an excess number of submillimeter luminous galaxies (SMGs), detected at 850{\micron}, in $z\sim$ 1 cluster fields and,
based on lensing considerations, argued that they were associated with the clusters rather than a background population. If the SMGs are indeed cluster members, they represent by far the most intense star bursts seen in clusters to date with rates of $\sim$100 M$_{\odot}$ yr$^{-1}$.  
Still, since none of the systems have been verified to lie at the cluster redshifts they could be background objects, perhaps lensed by the cluster potential.  

Lensing effects have been exploited by many groups  to overcome the blank-field confusion limitations and detect faint SMGs ($S_{850{\mu}m} <$ 2 mJy) at high redshift \citep{sma97,cow02,cha02}; but these surveys have employed lower redshift clusters  ($z \lesssim$ 0.5), since massive, relaxed clusters at low redshift are more efficient lenses than high-redshift clusters.  Recent results from the Red-Sequence Cluster Survey \cite[RCS;][]{gla05}, however, indicate that this is not always the case, and some higher redshift clusters appear to be excellent lenses.

 \citet{gla03} have found that the number of strong multiple optical  arcs for $z>$ 0.6 clusters  seen in the RCS  is larger than expected. This, along with the fact that the number of arcs in other medium-/high-redshift cluster surveys are several times larger than that expected from standard $\Lambda$CDM cosmology \citep{bart98,zar03,dal04}, leads to the interpretation that at least at high redshift, there may be a subset of rich clusters which have much higher lensing cross-sections than predicted \citep[but see][for an alternative explanation]{wam04}.  These ``super-lenses''  could be due, for example,  to an increase in cluster substructure with redshift, when clusters are observed in a younger and less relaxed state.

To examine both the issues of super-lensing and and possible excess SMGs in high redshift clusters, we  we have begun  a survey at 850$\mu$m  of high redshift (0.6 $< z <$ 1.1) galaxy clusters from the RCS with SCUBA on the JCMT.   The survey consists of two subsamples: five which show strong optical arcs and five control clusters of similar richness and redshift,  which do not.  Here we present the submillimeter data and follow-up imaging from  a single super lensing cluster field  RCSJ022434-0002.5 (hereafter RCS0224); the entire submm program, when complete, will be presented in a later paper (Webb et al., in preparation). This paper is organized as follows.  In \S 2 we outline the submillimeter and supplementary observations and data analysis;  \S3 presents the object catalogue; in \S 4 we discuss the properties of the counterparts to the submillimeter emission; \S 5 contains a discussion of the results.  A flat, $\Omega_\Lambda$=0.7 universe, with $H_\circ$=72 km/s/Mpc was assumed throughout.

\section{Observations and Data Reduction}

\subsection{RCS0224}

RCS0224  \citep{gla02,gla03} was discovered within the RCS, a 90-square degree optical imaging survey designed to find galaxy clusters up to $z=$ 1.4.  It is the most spectacular example of the RCS super lensing clusters, with two strong optical arcs originating from different background objects, one with $z$ = 4.9, and a third arc was discovered in a deep HST image. Subsequent spectroscopic follow-up has confirmed the cluster redshift to be $z=$ 0.773.

\subsection{Submillimeter Data}

\subsubsection{Observations and Reduction}

The cluster was simultaneously observed  at 450{\micron} and 850{\micron}  using the Submillimeter Common-User Bolometer Array (SCUBA) \citep{hol98} on the James Clerk Maxwell Telescope (JCMT) over two observing runs in 2001 and 2002. We employed the jiggle-map mode of SCUBA which fills in the under-sampled sky by stepping the secondary mirror through a 64-point pattern in each nod position throughout each 128-sec integration.  With AC-coupled bolometers, chopping and nodding is still required in the submillimeter for proper sky removal  and we chopped in RA with a 30{\arcsec} chop-throw.  Such a small chop-throw has two advantages: it optimizes sky removal by sampling nearby sky, and the off-source chop remains on the array, creating a negative-positive-negative beam pattern that is used to improve the extraction of  real sources.

The map consists of two pointings, centrally offset from each other by 20{\arcsec} in RA and Dec and roughly centered on the cluster center, as determined from the optical data.  The area is approxmately 6.3arcmin$^2$ though the depth varies with position in the map due to uneven coverage and differing bolometer sensitivities. A total integration time of 38.5ks was obtained  on the center of the map, but the effective integration time is less at the edges  where  the sky is not evenly sampled by the array and where pointings do not overlap. 

 The sky opacity  was monitored through {\it skydips} every 1-2 hours and was stable throughout  each individual night, ranging from ${\tau}_{850}\sim$ 0.1-0.32 throughout the two runs. Pointing stability was also closely monitored: pointing observations on near-by point-sources were performed every 1.5 hours or after a major slew of the telescope.   Calibration maps were acquired each night and calibration factors were found to be stable to within 10\% of the standard gain values at 850{\micron}. The calibration at 450{\micron} is stable to $\sim$ 30\%. 
 
The data reduction was complicated by the presence of a strong correlated noise signal which occurred at the jiggle-map frequency (16 seconds) (C. Borys personal communication, 2004). This signal appears in  the 2002 data and affects between 1/3 and 2/3 of the bolometers at any given time. We reduced its effect in the following way. 
After the standard   flat-fielding and extinction corrections, noise corrupted bolometers were identified through the presence of power in their fourier spectrum at the characteristic scale.  Residual sky-flux was removed from all the bolometers by subtracting the median sky level at each second, determined using only the non-corrupted bolometers.  Because the noise on the remaining corrupted bolometers is correlated it is possible to reduce its effect through the subtraction of this correlated signal. This was done through simple multiple linear regression techniques in which the expected noise signal in a given bolometer was predicted  using all other corrupted bolometers and then removed.   As a final step, noise spikes at $>$ 3$\sigma$ were iteratively removed from all bolometers and the bolometer time-streams were rebinned to produce the final map. The removal of the correlated noise from the corrupted bolometers results in a overall reduction of the noise in the final map of ${\sim}$ 30\%. A comparison of the  non-corrupted 2001 data with the corrupted  data from 2002, both corrected and non-corrected show that the presence of point sources in this map is robust, with no sources lost and no new sources introduced other than what is expected from an improved signal-to-noise ratio (S/N).  We note, however, that this is not the case for shallower surveys where this noise signal can lead to significant changes in the number of real and spurious sources detected \citep{saw04}.

\subsubsection{Source Detection}
Because of strongly varying noise properties (spatial and temporal), and the fact that one is generally working at the detection limit of the data,  deep imaging with SCUBA requires careful noise analysis,  to assign significance to each detection, and to minimize spurious sources.  To estimate the noise as a function of position on the image we employed two related techniques. First, following the  method  outlined in \citet{eal00} and \citet{web03a} we generated 500 Monte Carlo simulations of each bolometer time-stream, assuming Gaussian noise statistics, uncorrelated bolometers, and  without introducing any signal. The reduction procedure was repeated on these simulated data, resulting in a set of final rebinned maps with no sources but with  noise characteristics similar to the real data. A noise map is then simply the variance of the individual simulations. Second, using the real bolometer time-streams we employed a bootstrap algorithm to generate a variance map from 500 realizations.  The two techniques produced similar variance maps: the mean ratio between the two maps was $\sim$ 1.0, and varied by $<$20\% over the entire field.   The significance levels quoted in Table \ref{list} refer to the first method.  

Source extraction was performed using an iterative cleaning technique \citep{eal00}.  An initial list of source positions was produced by convolving the raw map with the beam template, which includes the off-beams produced by the chop.  This technique is similar to the template fitting used by other groups, \citep[e.g.,][]{sco02} and is advantageous over simple smoothing since it incorporates information (position and flux) from the two negative off-sources associated with each real source.    Beginning with an initial  list of $S/N> $ 3 detections, the map was iteratively cleaned  of sources in 10\% flux steps.   Using these cleaned results, for each source   all other sources were removed from the raw, unconvolved map,   and the isolated source was again convolved with the beam template.  This  provides improved flux densities, positions,  and detection significances for each source since  contamination from near-by confused sources is  reduced (see Figure 1 and \S3.1) 

The  450{\micron} observations, which are taken simultaneously by exploiting a dichroic beam splitter, suffer from a greatly increased sky opacity and a very unstable beam, and are therefore much more difficult to work with. Instead of generating a  separate source catalog  from the 450{\micron} data we simply searched for 450{\micron} emission from the previously detected 850{\micron} sources.   We adopted a search radius of 10{\arcsec} around each 850{\micron} position, allowing for a combination of the expected positional uncertainties at both wavelengths and we take as the flux measurement  the nearest 450{\micron} peak above 2.5$\sigma$ within this radius. 

\subsection{Supplementary Data}

 \subsubsection{Radio Imaging}
 
Approximately 5 hours of data were obtained on each of three days during December 2003 using the National Radio Astronomy Observatory's\footnote{NROA is operated by the Associated Universities Inc., under a cooperative agreement with the National Science Foundation.} (NRAO) Very Large Array in  B configuration, exploiting a pseudo-continuum correlator mode to minimize bandwidth smearing. 
Data were recorded every 5 seconds in 3.25-MHz channels, 28 in total, centered at 1.4\,GHz, taking both left-circular and right-circular polarizations. 0137+331 was used for flux calibration and 0239+42 (0.72\,Jy) was observed every hour for local phase/amplitude/bandpass calibration.

Initial calibration and mapping were performed using standard {\sc aips} tasks. The initial calibration was good on most baselines, though we were able to improve the fidelity of the final image by running a series of self-calibration/imaging tasks, with a model comprising the central 820$''$ $\times$ 820$''$ field plus 11 non-contiguous fields containing bright sources. 
The final image has a series of north-south stripes near the brightest sources, typical of equatorial fields, but in clean regions of the map the noise is around 15\,$\mu$Jy\,beam$^{-1}$, where the beam is near-circular with {\sc fwhm} $\sim$ 4.5$''$.

\subsubsection{Optical and Near-Infrared Imaging}
Multi-filter optical images of the field are available from various
sources.
The original discovery data from the RCS provide data in $z'$ and $R_c$
using the CFHT12K camera \citep{gla02}.
Additional images in $V$ and $B$ have been obtained, 
also with the CFHT12K, as
part of the four-color photometry follow-up of the CFHT part of the
RCS survey.  The details of the observations and photometric reduction techniques are provided in \citet{gla05}  for the $z'$ and $R_c$ data, and in \citet{hsi05} for the V and B images.

$J$ and $K'$-band images of the central ${\sim}$2${'\times}$2$'$ were obtained
using the PANIC camera \citep{mar04} on the Magellan Baade 6.5m telescope.
PANIC is a 1024$\times$1024 pixel near-IR camera with a pixel scale of
0.125$''$ per pixel.
Integrations totalling 54 min in $K'$ were obtained on 21/12/2003 and 52 min in $J$ on 24/10/2004, both under
moderately good seeing and photometric conditions.
Object finding and photometry in the $K'$ image were performed using 
the program PPP (Yee 1991).  
The $R_c-K'$ colors  were measured using identical angular aperture sizes
based on the magnitude growth curve \citep[see][for details]{yee91}.   $(J-K')$ colors, discussed in \S5, were measured for the two brightest SMGs in a similar manner.

\section{Submillimeter Results}

\subsection{The  Detected Sources}

\begin{figure*}
\epsscale{1.1}
\plottwo{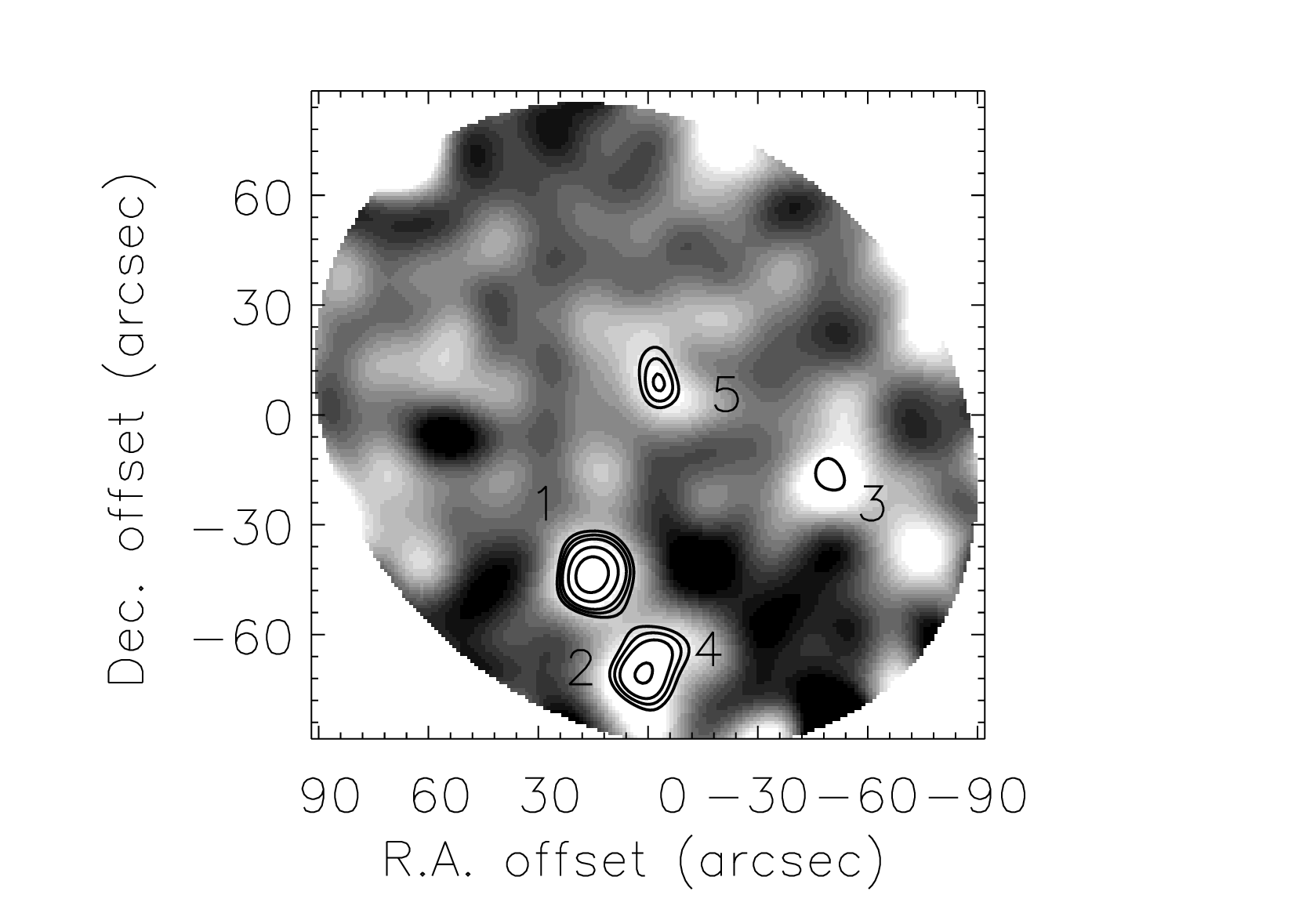}{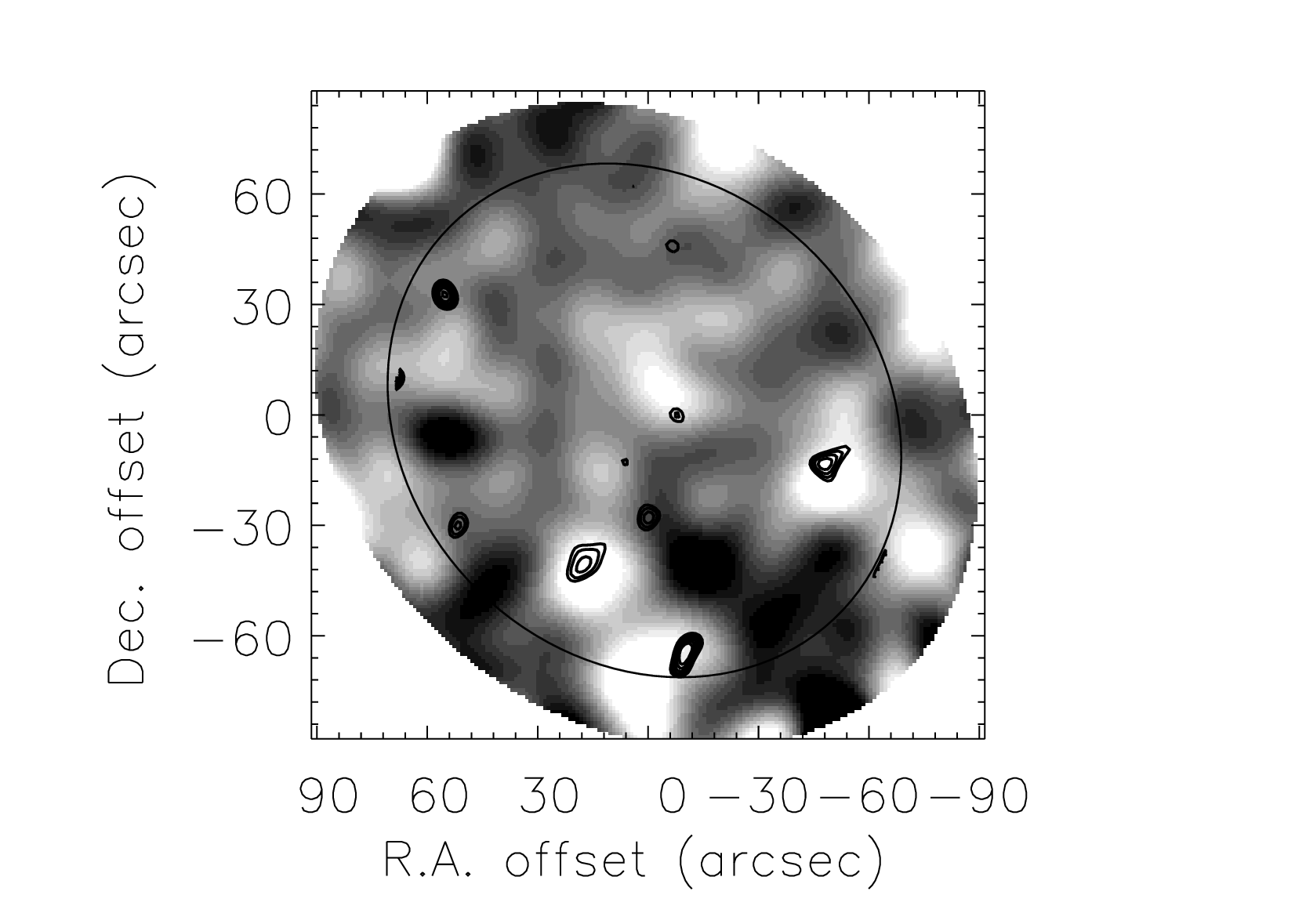}
\caption{Left: The 850{\micron} map of RCS0224, smoothed with a 14{\arcsec} Gaussian, with white indicating positive flux. Overlaid are the  S/N contours (before cleaning) in 0.5 steps, beginning with 3.0${\sigma}$ and 1.0 steps above 4$\sigma$. The five sources listed in Table \ref{list} are marked. Sources 2 and 4 are confused and separate into two sources during the 850$\mu$m cleaning process. This more easily apparent in the 450$\mu$m map in which source 4 has been detected. Right: The greyscale 850{\micron} map of RCS0224  with the 450{\micron} flux contours overlaid. The contours have been  smoothed with an 8{\arcsec} Gaussian and are shown in steps of 1mJy starting at 12mJy.  Note that the 450{\micron} FOV (denoted by the ellipse) is smaller than at 850{\micron} such that the outer $\sim$20{\arcsec} of the 850{\micron} map does not have useful 450{\micron} coverage, including the southern-most 850{\micron}.  \label{map}}
\end{figure*}

The 850{\micron} map is shown in Figure \ref{map} and the detected sources are listed in Table \ref{list}.  We detect five SMGs at 850{\micron} above $S/N>$ 3.0.  
Because the use of chopping to remove the sky level results in a map with a mean flux level of zero, the noise will be symmetric around zero and and the  number of spurious sources on a SCUBA map may be estimated through the number of sources detected in an inverted map. After first removing all positive sources and their associated negative off-sources, we searched for negative sources at $>$ 3.0 $\sigma$ level on the inverted map. Only one negative source was found, at 3.0$\sigma$ near the northern edge of the map, indicating  our source list is reliable at this level.

To quantify the uncertainties in the submillimeter position and flux measurement we placed and recovered point sources  of varying flux levels and positions into the image.  Working with the real map will produce more realistic results than working with the simulated data, since the effects of confusion are included through the presence of the unresolved background. In order to avoid over-estimating the effects of confusion the five bright real sources (i.e., S/N $>$ 3) were removed from the map before this analysis. Given our small field size and the high number of sources the addition of even a single object leads to  an unrealistically large source density. On the other hand, removing all the detected sources will lead to an underestimate of the positional uncertainties, since only confusion with sources below the confusion limit is considered.

The 95th percentiles for the positional offsets for S/N of 3.5$\sigma$, 4.5$\sigma$, 5.5$\sigma$, and 6.5$\sigma$ are 9.2{\arcsec}, 8.4{\arcsec}, 7.6{\arcsec}, and 7.6{\arcsec} respectively.  The distributions in each bin are all centered at $\sim$ 4.5{\arcsec} and skewed strongly to larger offsets, particularly in the lowest S/N bins.  We note that these  offsets are in good agreement with \citet{ivi05} who find a positional accuracy (1$\sigma$) of 1.5$\times$FWHM/SNR for 5$\sigma$ sources, determined from the distribution of offsets from the objects to radio positions.   The error on the flux measurement may be estimated from the signal-to-noise map, or through a comparison of the input flux to recovered flux in the simulations described above (more correctly, from the output flux to the range in input fluxes which produce such a measured flux).  It was found that these two methods compared well, indicating that the Monte Carlo simulations reasonably  reproduce the noise statistics of the data. 

In Figure \ref{map}(b) we show the 450{\micron} flux contours, overlaid on the 850{\micron} map, and a  correlation between the two wavelengths is apparent. Four of the five 850{\micron} sources have usable 450{\micron} coverage (the southern-most 850$\mu$m source lies in the unreliable edge region of the 450{\micron} map). Three of these are detected at 450$\mu$m with  $\geq$ 3.0$\sigma$, and the fourth with 2.5$\sigma$.  Using the 450{\micron} number density on the map at these levels of significance we can estimate the probability that a 450{\micron} detection is randomly aligned within 10{\arcsec} of an 850{\micron} source.  For the 450{\micron} sources with S/N$\geq$ 3 this probabilty is 8\%, while the single S/N = 2.5$\sigma$ has a probability of 16\%.  Given the actual measured offsets (which, barring one, are  $\lesssim$ 6{\arcsec}), the probability of chance alignments for all four sources is $<$ 6\%.  Thus, for these four sources we would expect less than one chance alignment and regard the 450{\micron} detections as reliable counterparts to the 850$\mu$m emission.  The detection of these four objects in both submillimeter bands  is strong evidence that they are real and not spurious detections.

In principle, because of the smaller beam size, the 450{\micron} detection should provide a better measure of the true source position than at 850{\micron}; in practice, the poorer quality of the 450{\micron} data does not make this so.  Simulations of the positional uncertainty at 450{\micron}, similar to those discussed above for 850{\micron}, indicate that at $\lesssim$ 4$\sigma$ (the significance of these detections) the median positional uncertainty is roughly equal to that at 850{\micron} but with higher dispersion and skewed heavily to large offsets (the 95\% percentile is $\sim$ 13{\arcsec}). 

\begin{deluxetable*}{ccccccccc}
\tabletypesize{\scriptsize}
\tablewidth{0pt}
\tablecaption{Positions and Flux Densities of Submillimeter Sources \label{list}}
\tablehead{
\colhead{Name} & \colhead{R.A.} & \colhead{Dec.} & \colhead{$S_{850{\mu}m}$} & \colhead{850$\mu$m} &  \colhead{$S_{450{\mu}m}$} & \colhead{450{\micron} } & \colhead{offset: 850{\micron} and } & \colhead{lensing} \\
\colhead{} & \colhead{(J2000)} & \colhead{(J2000)} & \colhead{(mJy)} & \colhead{ S/N} & \colhead{(mJy)} & \colhead{S/N} & \colhead{450{\micron} pos.(\arcsec)} & \colhead{magnification\tablenotemark{a}}
}
\startdata
SMM-RCS0224.1 & 02:24:34.29  & -00:03:28 & 6.4 & 7.0  & 14.4   & 3.5 &  5.2  & 1.10-1.17  \\
SMM-RCS0224.2 & 02:24:33.56 & -00:03:58 & 5.3 & 4.6 & $>$ 66\tablenotemark{b} &   ... &  ... & 1.06-1.10 \\
SMM-RCS0224.3 & 02:24:29.79  & -00:03:01 & 4.0  & 3.3 &  15.8  & 3.0 & 6.4   & 1.05-1.09 \\
SMM-RCS0224.4 & 02:24:32.69 & -00:03:46 & 3.2 &  3.1 &  18.7   & 2.5 & 1.6   & 1.06-1.11 \\
SMM-RCS0224.5 & 02:24:33.00 & -00:03:35 & 3.2 & 4.1 &  13.2  & 3.0 & 9.0   & 1.61-2.59 \\
\enddata
\tablenotetext{a}{For all sources but SMM-RCS0224.4 the lensing magnification was determined using the radio position (see Table \ref{radio}). Two possible magnifications are listed: the first assumes a source redshift of $z=$ 1.5 and the second of $z=$ 4.0.  Note that this is simply the lensing calculated at the source position and none of  these sources are confirmed as lensed objects.}
\tablenotetext{b}{A 3$\sigma$ upper-limit on the flux; the object lies at the edge of the usable 450$\mu$m field.} 
\end{deluxetable*}

\begin{deluxetable*}{ccccccccc}
\tabletypesize{\scriptsize}
\tablewidth{0pt}
\tablecaption{The Counterparts \label{radio}}
\tablehead{
\colhead{Name} & \colhead{ R.A.} & \colhead{ Dec}  & \colhead{1.4GHz flux} & \colhead{offset from }  & \colhead{$z_{est}$\tablenotemark{a}} & \colhead{K$'$} & \colhead{$(R_C-K')$} & \colhead{$z_{phot}$\tablenotemark{c}} \\
\colhead{} & \colhead{1.4GHz} & \colhead{1.4GHz} & \colhead{ ($\mu$Jy)} & \colhead{850{\micron}(\arcsec)} & \colhead{} & \colhead{} & \colhead{} & \colhead{} }
\startdata
SMM-RCS0224.1 & 02:24:33.815 & -00:03:33.00 & 1652 $\pm$ 41 & 8.7 & $<$ 0 \tablenotemark{b} & 16.92 $\pm$ 0.01 & 4.63 $\pm$ 0.02  & 0.66 $\pm$ 0.11 \\
SMM-RCS0224.2 & 02:24:33.332 & -00:04:05.53 & 68.4 $\pm$ 29 & 8.3\tablenotemark{d} & 1.9 $\pm$ 1.0 & 18.22 $\pm $ 0.04 & 4.11 $\pm$ 0.05  & 0.72 $\pm$ 0.08  \\
SMM-RCS0224.3 & 02:24:29.972 & -00:03:04.10 & 97.4 $\pm$ 29 & 4.1 & 0.85 $\pm$ 1.0 & 20.81 $\pm$ 0.14 & $\sim$6.1 $\pm$ 0.7{ } & ...  \\
SMM-RCS0224.4 & ... & ... & $<$ 48\tablenotemark{e} & ... & $>$ 1.6\tablenotemark{e}  & ...& ...  & ... \\
SMM-RCS0224.5 & 02:24:32.867 & -00:02:31.34 & 94.3 $\pm$ 28 & 4.1  & 0.65 $\pm$ 1.0 & 19.38 $\pm$ 0.06 & 4.6 $\pm$ 0.12  & ...\\
\enddata 
\tablenotetext{a}{Using the empirical 1.4GHz-850$\mu$m relation for high redshift submm-selected galaxies from \citet{cha05}. All estimates have uncertainties of $\Delta z\sim$ 1. }
\tablenotetext{b}{The substantial radio emission of this source, and the compact nature of the optical counterpart leads us to conclude that this object contains an AGN. As discussed in the text, this will result in an underestimate of the redshift and in this case a completely non-physical estimate of $z<0$.   }
\tablenotetext{c} {The photometric redshift is determined using the $z'R_cVB$ empirical training set technique presented in \citet{hsi05}. Sources without a photometric redshift have not been detected in enough filters to enable an estimate.}
\tablenotetext{d}{The large positional offset could be due, in part, to blending with SMM-RCS0224.4}
\tablenotetext{e}{A 3$\sigma$ upper limit on the flux.}
\end{deluxetable*}

\subsection{Gravitational Lensing}

The magnifications were computed using a model that was derived to
reproduce the strong lensing features observed for the cluster.  The
model consists of a smooth cluster component which was modeled as an
isothermal ellipsoid with a core. The free model parameters are the
central velocity dispersion, core radius, ellipticity, position angle
and the position of smooth halo. In addition we included truncated
isothermal spheres at the positions of galaxies on the cluster
$R_C-z'$ color-magnitude relation. For these we assumed a Faber-Jackson
relation and used the velocity dispersion of an L$^{\star}$ galaxy as a free paramter in
the model.

We used  deep optical HST observations\footnote{Based on observations made with the NASA/ESA Hubble Space Telescope, obtained at the Space Telescope Science Institute, which is operated by the Association of Universities for Research in Astronomy, Inc., under NASA contract NAS 5-26555. These observations are associated with program \# 9135} to model the lensing but 
found that the model cannot reproduce all strong lensing feature to
great accuracy.
The L$^{\star}$ velocity is not well constrained, but the best fit value of 180
km/s is quite reasonable based on other studies (e.g., Hoekstra et
al. 2000). The models prefer a small core radius.
Despite the problems of reproducing all detailed lensing features, we
found that the results for the mass, ellipticity and position angle of
the smooth cluster component are robust.  This is not surprising
because these are well constrained by the high redshift arc system,
for which we have confirmed the counter image spectroscopically. The orientation based on
the strong lensing model agrees well with that of the X-ray emission \citep{hic04}.
The resulting mass estimate is also in good agreement with that determined
from a weak lensing analysis of the HST image.

Given the fact that most of the SMGs are found at relatively large
distances from the cluster center, the derived strong lensing model
is sufficiently accurate to derive their magnifications. 
In Table \ref{list} we list the lensing magnifications these objects would experience assuming they lie behind the cluster.  The magnifications are calculated assuming source plane redshifts of $z=$ 1.5 - 4.0 though in Table \ref{list} we list only the two extremes. Four of the five of these objects lie at large angular distances from the cluster center and therefore suffer very small magnifications of $\sim$ 1.1 for all plausible redshifts. Even source SMM-RCS0224.5 which lies close to the cluster center  has a magnification of only 2.3 in the most extreme case.  In conclusion, simple but accurate analysis shows that on average we do not expect the SMGs to be significantly magnified.

\section{The Radio and NIR Counterparts and Redshift Estimates}

\begin{figure*}
\plottwo{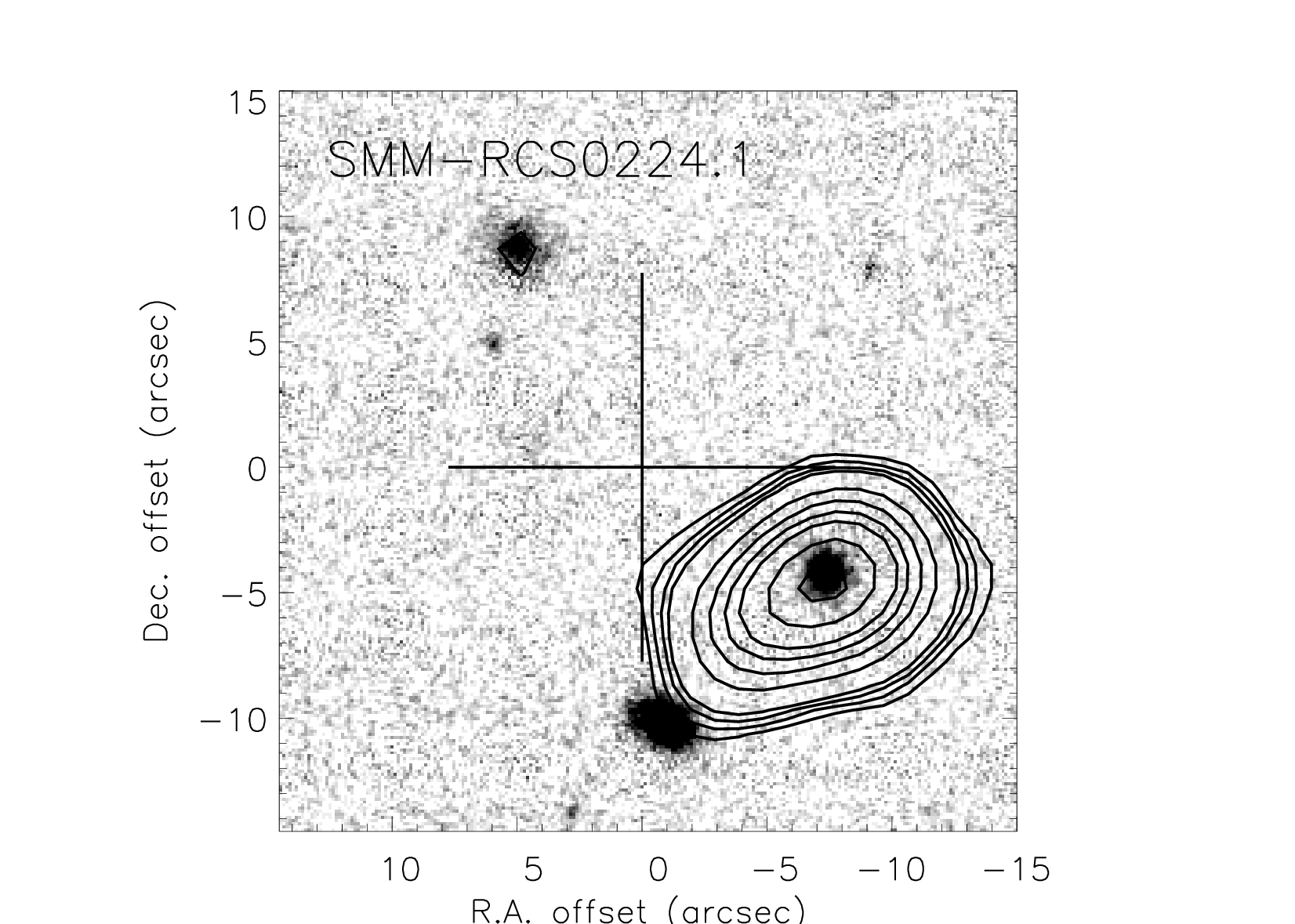}{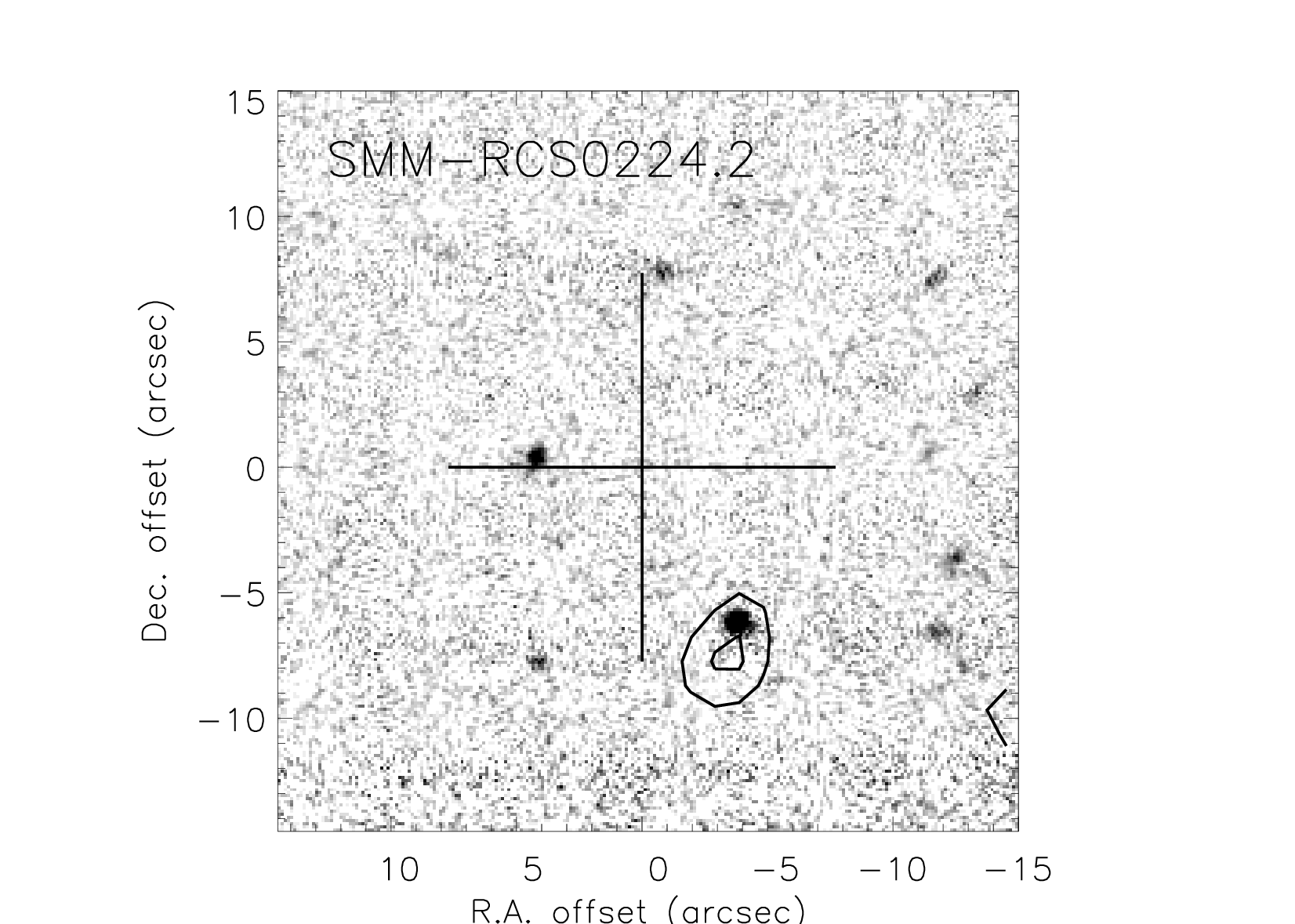}
\plottwo{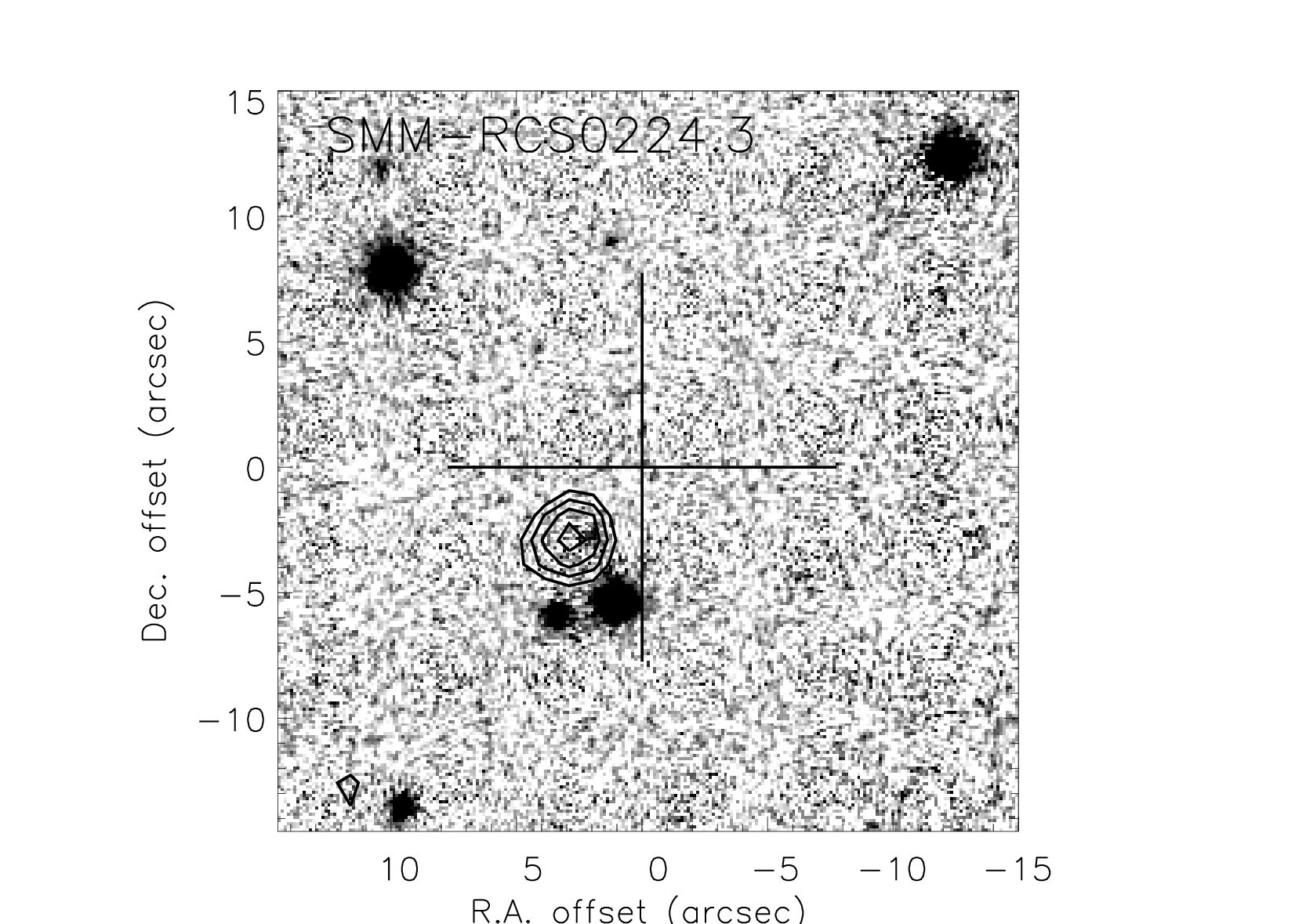}{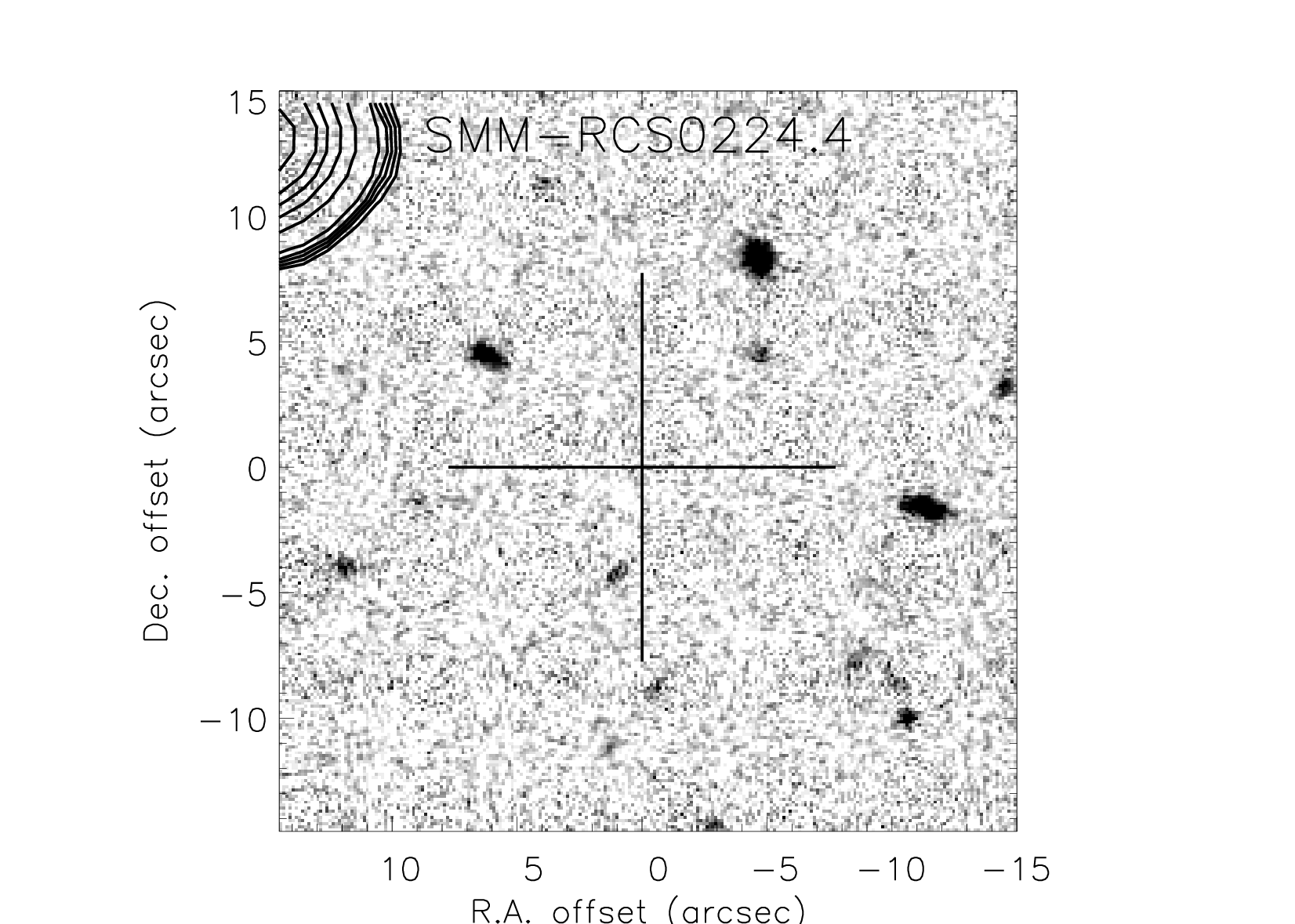}
\epsscale{0.5}
\plotone{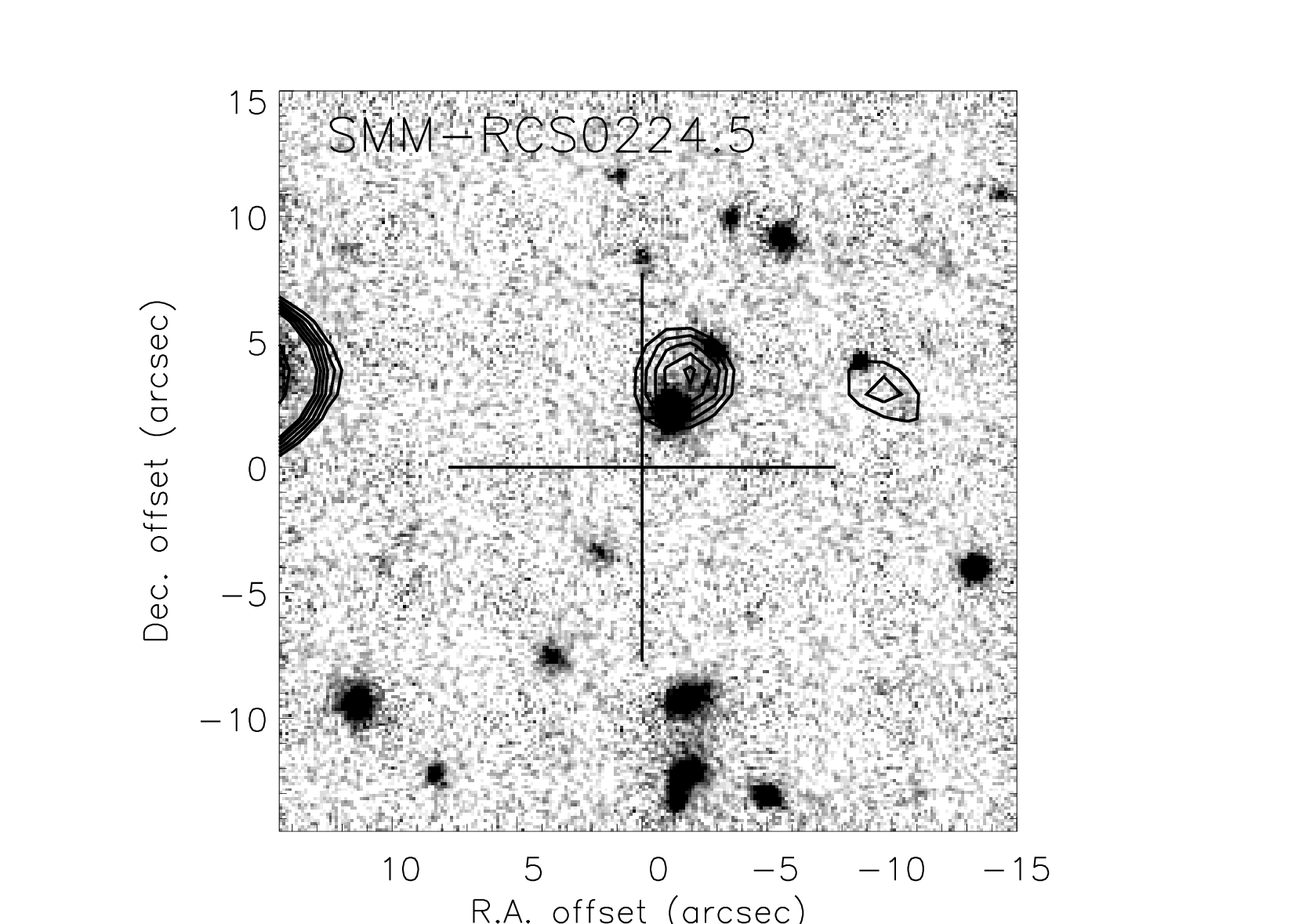}
\epsscale{1.0}
\caption{$K'$ images of the region surrounding each SMG. The images are centered on the SMG position with the approximate positional error indicated by a cross (8\arcsec, $\sim$95\%). Overlaid are the 1.4GHz contours showing the radio detections.  Source SMM-RCS0224.4 does not have a radio identification. The offset between the radio and optical positions for source SMM-RCS0224.2 is likely due to the low S/N of this object in the radio. East is to the left and North is up. 1.4GHz contours start at 40${\mu}$Jy and increase in 20${\mu}$Jy steps up to 100${\mu}$Jy, and then in 100${\mu}$Jy steps.\label{ids}}
\end{figure*}

\begin{figure*}[h]
\vspace*{5cm}
\plottwo{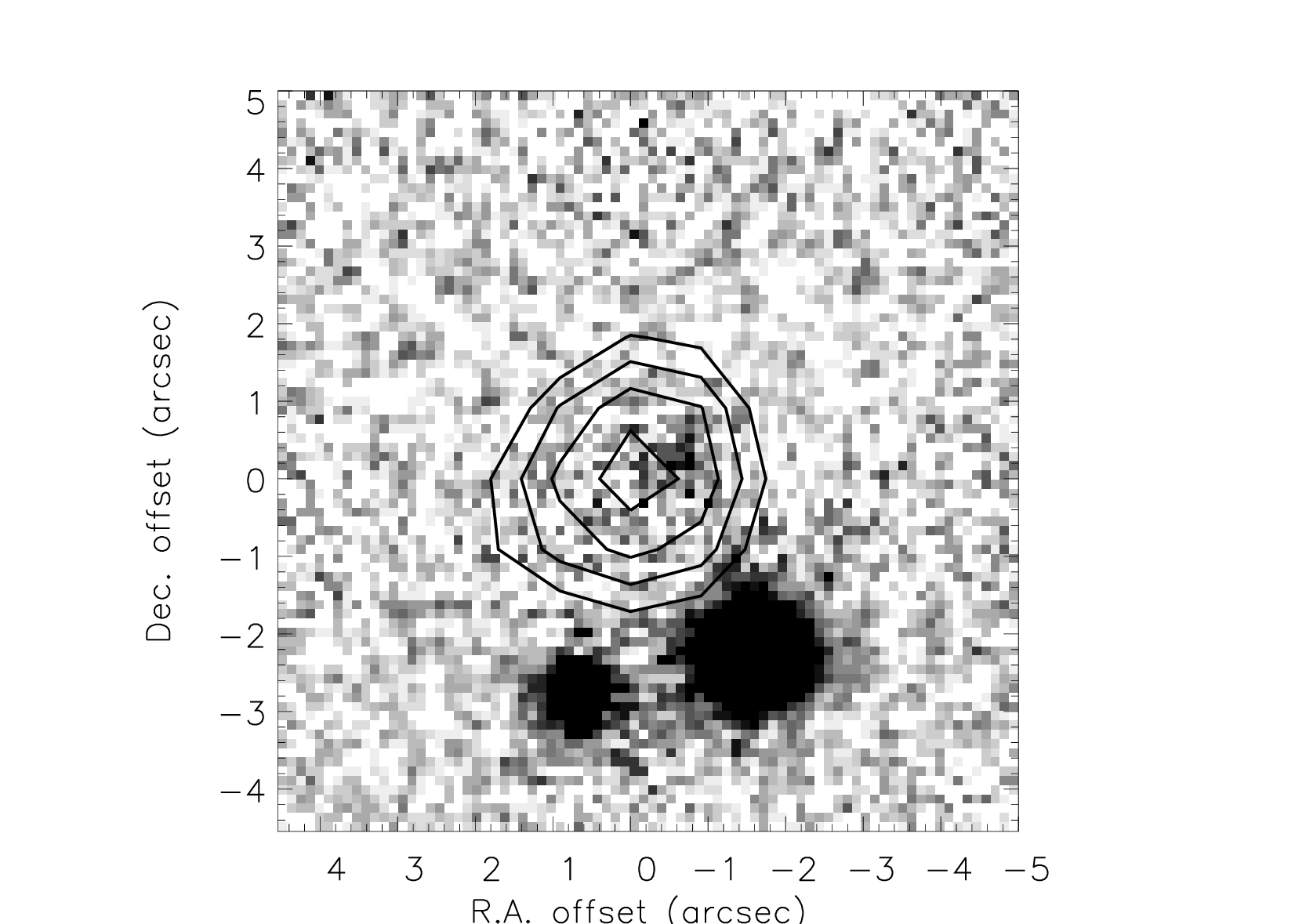}{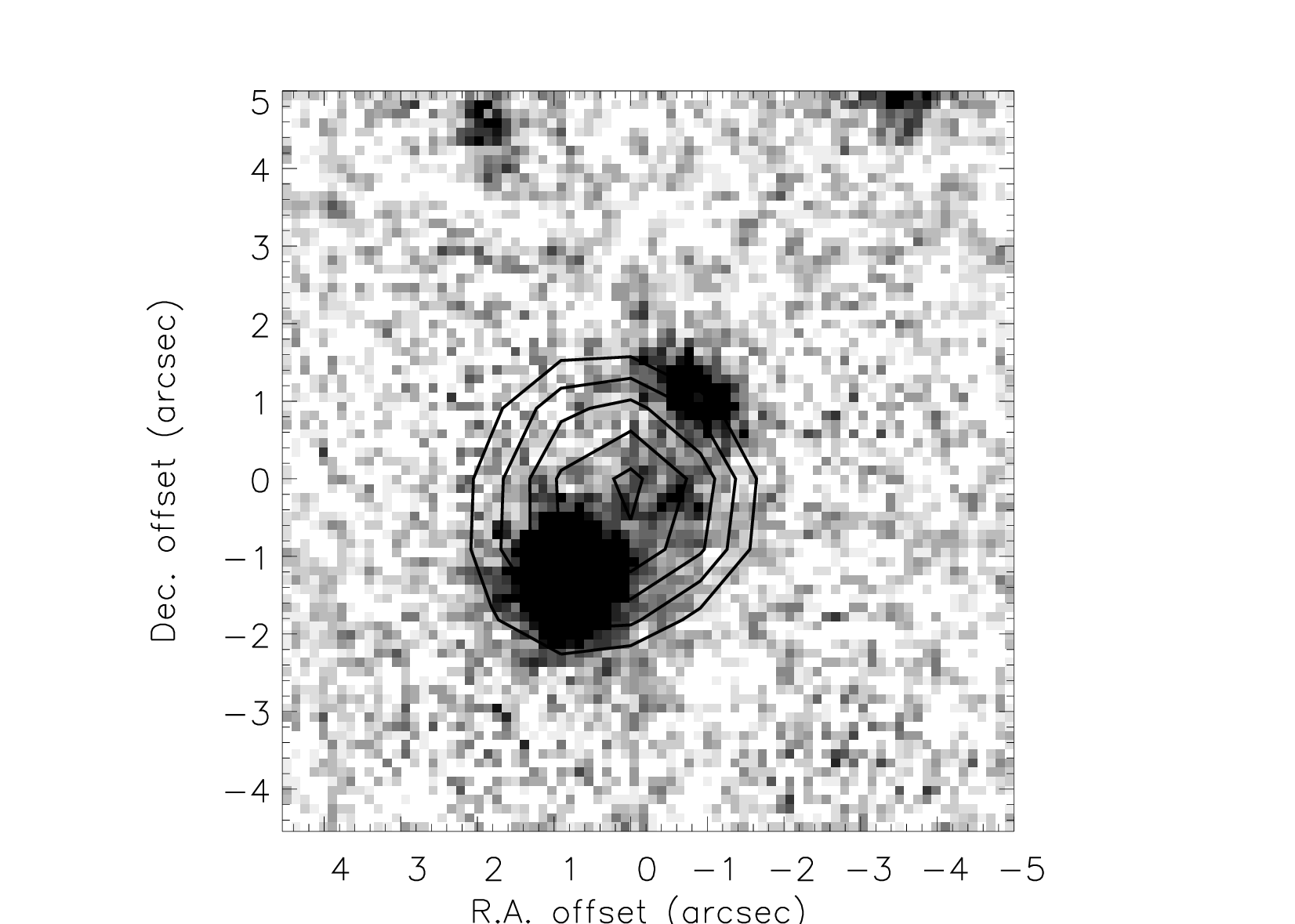}
\caption{Enlarged  $K$-band images and radio contour maps of  Fig.\ref{ids} for SMM-RCS0224.3 (left) and  SMM-RCS0224.5 (right), showing the faint $K$-band identifications and the close neighbors.  \label{ids2}}
\end{figure*}

To identify counterparts to the SMGs we employed the empirical correlation between the radio and far-infrared flux \citep{con92} observed in the local universe.  Using the deep 1.4-GHz maps we searched for a radio detection within the submillimeter error radius (\S3.1).    Statistically this is also one of the least ambiguous methods of identifying counterparts since the number density of radio sources on the sky is low enough that chance alignments between submillimeter and radio galaxies are almost negligible. In Figures \ref{ids} and \ref{ids2}  we show the  $K'$-images surrounding each submm detection, with 1.4-GHz flux contours overlaid.   Four of the five 850$\micron$ sources have radio detections within the submillimeter search radius (\S 3.1) and all of these four have $K'$ counterparts.
The fifth source, SMM-RCS0224.4, does not have a secure counterpart at radio or optical/near-infrared wavelengths.

The $z'R_cVB$ four-filter photometry data were used to generate photometric redshifts using an empirical training set technique, providing photometric redshifts with typical accuracy of $\sim10$\% for galaxies with $R_c \lesssim$ 23.0 (for details, see Hsieh et al.~2005).  Two of the four radio-detected objects (SMM-RCS0224.1 and SMM-RCS0224.2) have sufficiently accurate optical
measurements to allow photometric redshift estimates. We were not able to obtain phtometric redshifts for the other two sources with radio counterparts. The sources SMM-RCS0224.3 is undetected in all our optical bands while SMM-RCS0224.5 is detected only marginally in the $R_c$ image.  The redshift estimates are listed in Table \ref{radio} along with the  $K'$ mag and $R_C-K'$ colors.

\begin{figure*}
\epsscale{0.8}
\plotone{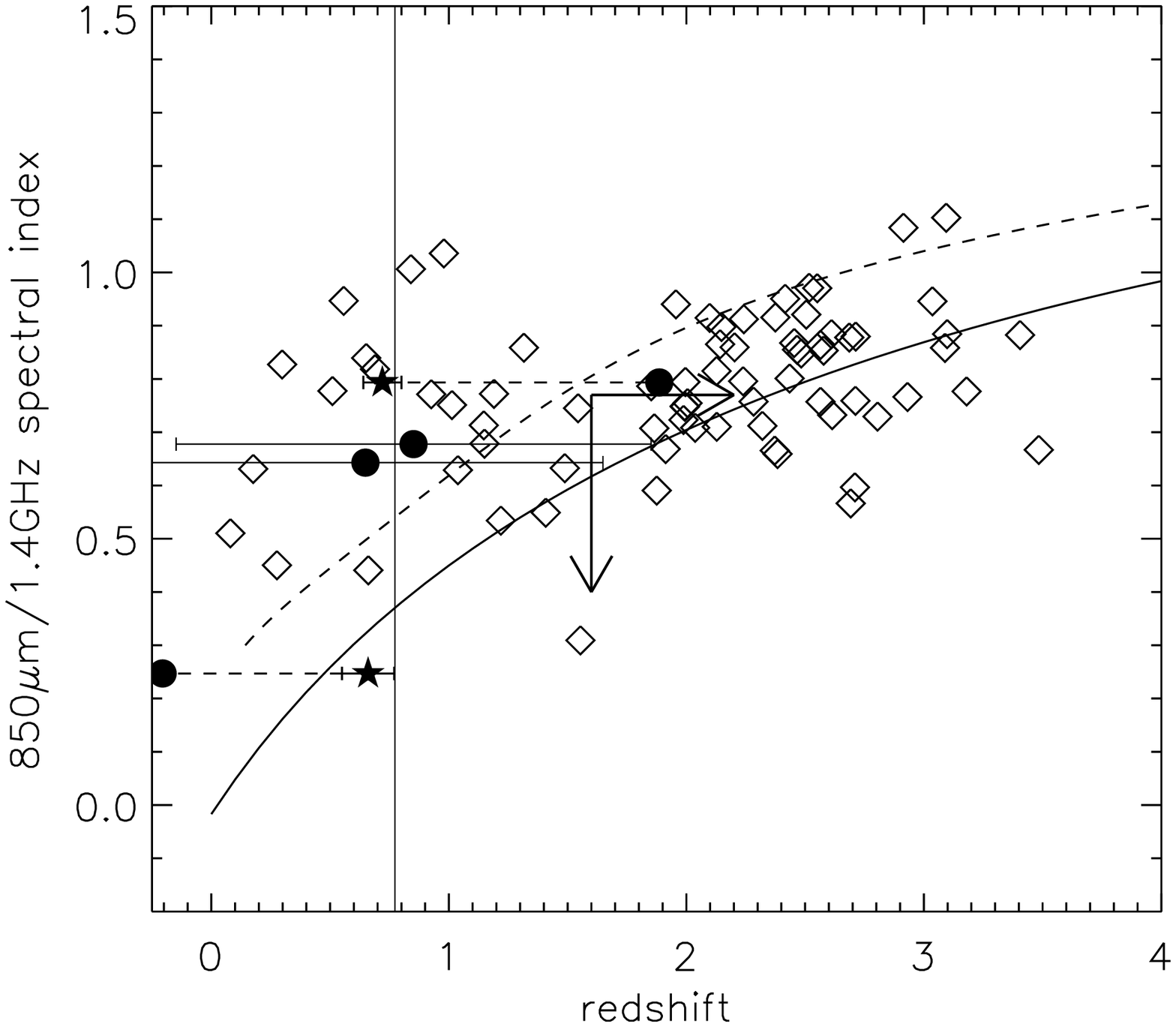}
\caption{The radio-submm spectral index $\alpha = 0.42 log (S_{850{\mu}m}/S_{1.4-GHz})$ as a function of redshift.  The open diamonds correspond to submm-selected galaxies with 1.4GHz counterparts and spectroscopic redshifts \citep{cha05}.  Also shown are two relations empirically determined using low-redshift starburst galaxies: solid line \citet{yun02} and  dashed line \citet{dun00}.  The cluster redshift is denoted by the vertical solid line.  The radio-submm redshift estimates for the four RCS0224 sources  with radio detections are shown by the solid circles; these redshifts have been determined from the average relation of the submm-selected population (i.e., the open diamonds).  The optical photometric redshift estimate of sources SMM-RCS0224.1 and SMM-RCS0224.2, are shown by the solid stars and are connected to their corresponding radio estimate by dashed lines. We use the scatter of the \citet{cha05} data to estimate the uncertainties of the radio-submm redshift redshift estimate which we plot as error bars for SMM-RCS0224.3 and SMM-RCS0224.5. For clarity, we have not plotted the errors for SMM-RCS0224.1 and SMM-RCS0224.2 which are similar to the other two sources.  Note that the  $\alpha$ value for SMM-RCS0224.1 implies a negative redshift which is likely due to  the radio flux of this source being highly contaminated by emission from an AGN (see text).  Source SMM-RCS0224.4 does not have a radio detection and the upper-limit on $\alpha$, and the corresponding redshift limit are denoted by the errors. This limit carries the same uncertainty as the measured ratios.  \label{redshift}}
\end{figure*}

A detection at radio wavelengths has the added advantage that it  provides a redshift constraint, from the radio-submm spectral index \citep{dun00,yun02}.  This method draws on the tight correlation between FIR and radio flux for low redshift star forming galaxies and the steep slope of the submm side of the thermal dust spectrum \citep[e.g.,][]{eal00,ivi02}.  
 There are, however,  a number of substantial uncertainties  in this technique including the assumed temperature of the dust and  contamination of the radio flux by an AGN.  
Until recently  the small number of SMGs with spectroscopically determined  redshifts made it difficult to verify the extension of this relationship to the high-redshift 850$\mu$m selected population.    In  Fig.~\ref{redshift} we show the 1.4-GHz-850$\mu$m spectral index for the  SMGs with spectroscopic redshifts  from \citet{cha05} with two empirically determined relations from local starbursts overlaid. It is clear the SMG spectral indices  are not well described by the empirical low-redshift relations at $z\lesssim 1.4$ and therefore we adopt the average relation from the high redshift SMGs \citep{cha05} to  estimate redshifts for the four radio detected sources in the RCS0224 field (and a lower-limit for the fifth). These are listed in Table 2 and shown in Fig.~\ref{redshift}. We note that all have a substantial uncertainty of $\Delta z\sim$1.

Source SMM-RCS0224.1 is particularly interesting. It is the brightest submm detection in the RCS0224 field and is identified with a bright and extended radio source with a large positional offset of 8.7{\arcsec}. Since the positional error analysis outlined in \S 3.1 indicates that roughly  5\% of the detected sources should be recovered at $>$8{\arcsec} from their true position it is worrying  that this is one of two radio identifications with $>$8{\arcsec} in this field. However, given the extreme observed luminosity of the radio source, it is statistically very unlikely that the radio source and submm source are not related \citep[0.1\% assuming the 1.4GHz counts of ][]{bon03} and we reiterate that the positional uncertainty could be underestimated (\S3.1)

The radio counterpart is asymetrically extended over 7.0\arcsec$\times$4.8{\arcsec}, and centered on a highly compact, though resolved, galaxy in the optical and NIR images. The 450{\micron} emission (which should be resolved given the smaller 450{\micron} beam) shows hints of extension along the same position angle (see Figure \ref{map}).  The extreme radio flux of the source indicates the presence of an AGN: assuming the cluster redshift and a radio spectral index of $S{\propto}{\nu}^{-0.7}$ yields a rest frame luminosity of 1.9$\times$10$^{24}$ W Hz$^{-1}$, well within the radio power regime dominated by AGN \citep{con02}. The asymmetric extended emission is suggestive of the morphology of a head-tail radio source, and specifically a Narrow Angle Tail galaxy \citep[e.g.,][]{sij98}: a single, or parallel radio jet, bent by bulk flows in the intercluster medium. The presence of a strong AGN can introduce uncertainties to the photometric redshift, however,  this is unlikely to be significant in this case  since the object is resolved optically.

\section{Discussion}

\subsection{SMG Source Counts}
A main goal of our submillimeter survey of RCS clusters is to determine whether some, or possibly all, $z\sim$1 galaxy cluster fields show an excess number of SMGs compared to that expected from the blank-field counts and simple lensing estimates.  If so, this could be due either to enhanced gravitational lensing 
as explained in the \S 1 or to source count contamination by submillimeter-bright cluster members. 

\begin{figure*}
\epsscale{1.0}
\plotone{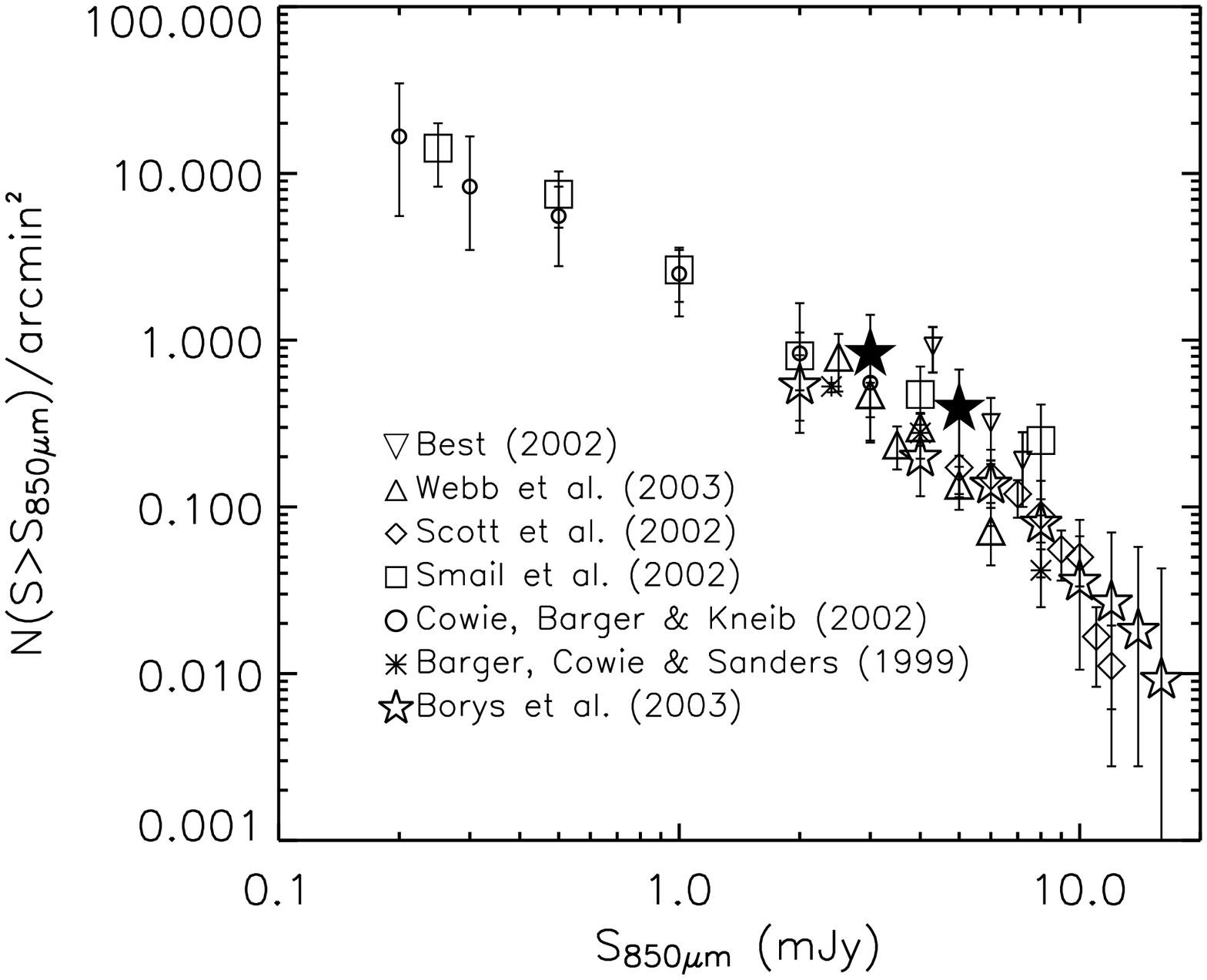}
\caption{The cumulative source counts for RCS0224 (large solid stars), assuming none of the detected objects have been lensed by the cluster. Also shown are the source counts determined from blank field surveys and cluster lensing surveys, corrected for lensing (Smail et al.~and Cowie et al.) and not corrected (Best). Note that 3/4 of the cluster surveys (Smail, Best, this work) lie systematically above the blank field counts at the 1$\sigma$ level. As mentioned in the text, this may be an indication of unrecognized submillimeter-luminous cluster members enhancing the counts, even at low redshift. \label{counts}}
\end{figure*}

 In Figure \ref{counts} we show the cumulative source counts toward RCS0224,  with no correction made for lensing effects, which we calculate to be negligible, but accounting for the varying depths within the submillimeter map.
The results from this single cluster field are consistent with the blank field counts and indicate an excess of sources due to either lensing or submillimeter-luminous cluster members only at the 1$\sigma$ level.
Interestingly, the source counts are  similar to those from other cluster surveys of \citet{sma02} and \citet{bes02b}  which also lie systematically above the blank-field counts at at least the 1$\sigma$ level  (the exception to this is  \citet{cow02} which is also a cluster program).  The Smail et al.~counts do not include two cluster members (cD galaxies)  and have been corrected for lensing which can be substantial for low redshift clusters, whereas the higher redshift cluster fields of Best (and RCS0224) suffer a significantly lower average lensing magnification (\S 3.2).  While the offset from the blank-field is small for each survey, taken together, these results suggest that source counts in cluster fields (at all redshifts)  are enhanced, possibily due to cluster member contamination. This  has been noted previously by \citet{bor03}. The  full RCS submillimeter  survey will increase the statistics on cluster field counts, in particular at higher redshifts where the enhancement may be more pronounced, due to younger clusters and hierarchical processes (discussed below).

\subsection{Dusty Star Formation in RCS0224?}
Although the RCS0224 cumulate source counts are consistent with the blank-field measurements, this in itself is not unambiguous evidence that all five  SMGs  are background systems; however, taken with our simple but adequate mass model (\S 3.2), it does indicates that none of these systems are substantially lensed.   Because of the scatter in the blank field source counts and the small area surveyed here, a scenario where some (but not all) of the SMGs are cluster members easily remains in-line with the expected source counts.  

The strongest evidence that some of the SMGs presented here are cluster galaxies  are the optical photometric redshifts of two of the radio-detected SMGs, SMM-RCS0224.1 and SMM-RCS0224.2, which are in excellent agreement with the cluster redshift (Table \ref{radio}).  The uncertainties on the optical photometric redshifts (e.g., $ \Delta z\sim$ 0.1) are substantially lower than for the radio-submm redshift estimate ($\Delta  z\sim$1.0); for this reason whilst the radio-submm redshift estimates are discussed below we give them much less weight than the optical photometric redshifts for these two galaxies. 

The near-infrared counterparts of SMM-RCS0224.1 and SMM-RCS0224.2 are shown in Fig.~\ref{ids2} and have colours and magnitudes ($K' \lesssim 18$; see Table \ref{radio}) suggestive of low-redshift objects. Their $R_c - K'$ colours are consistent with both the cluster red-sequence at this redshift, $(R_{702}-K)$ = 4.7 \citep{kod98}, as well as  dusty star-bursts and radio galaxies at $z\sim1$ \citep{sma04,wad00}.   The near-infrared color of SMM-RCS0224.1 is quite blue with $(J-K)$=1.1, similar to other  $z\sim 1 $ radio galaxies (which this source cleary is with $S_{1.4-GHz}$ = 1.6mJy), and possibly indicating substantial H$\alpha$ and NeII contamination of the $J$-band flux.   Source SMM-RCS0224.2  has $(J-K)=2.2$, which if at $z\sim 0.7$ (the photometric redshift and cluster redshift) suggests a dusty starburst, rather than an early-type galaxy \citep{poz00}. 

The remaining two radio-detected SMGs are much fainter in the optical bands and thus do not have sufficient detections  for phototmetric redshifts; however, the radio identification allows us to examine the $R_c-K'$ colors (our $J$ imaging is too shallow for $J-K$).  Source SMM-RCS0224.3 is very faint ($K=20.8$) with an extremely red color of at least $(R_c-K')\sim 6.1$.  This is consistent with a high-redshift $z\gtrsim 1$ galaxy and indeed with the counterparts of the $z\gtrsim 2$ blank-field SMGs which are often faint and red. Source SMM-RCS0224.5 however, has a more moderate $R_c-K'$ color similar to SMM-RCS0224.1 and SMM-RCS0224.2.

Redshifts estimated from radio-submm spectral indices  have substantial uncertainties ($\Delta z\sim$ 1) and provide much poorer redshift constraints than the photometric redshifts. Nevertheless,  since this is a frequently applied technique in the absence of spectroscopy, we discuss such estimates here.   This method implies an unphysical redshift of $z<0$ for SMM-RCS0224.1 which is almost certainly due to the AGN contribution to the radio flux, in addition to the usual uncertainties.  Two sources, SMM-RCS0224.3 and SMM-RCS0224.5 have measured ratios which are consistent with $z<$ 1 objects. As discussed above, however, the optical/NIR properties of SMM-RCS0224.3 do not support cluster membership but rather are similar to the background SMG population. Source SMM-RCS0224.5 on the other hand has a more moderate color and thus, while not ruling out a background source, is in general agreement with the radio estimate.   The radio-to-submm redshift estimate of SMM-RCS0224.2 is  $z$ = 1.9, substantially higher than the optical/NIR estimate and  while the radio-submm estimate is uncertain this raises the interesting possiblitity that the radio-submm galaxy  and the optical counterpart may represent different galaxies, as there is a small offset ($\sim$ 1 arcsec) between the radio and optical positions.   

In principle the 450$\micron$-850$\mu$m flux ratios may also be used to isloate low-redshift SMGs but the large uncertainties on the 450$\micron$ flux measurement and the flatness of the observed ratio with redshift out to $z\sim 2$ \citep[e.g.][]{web03a} makes this method of little use in our case.  Three of the four 450$\micron$ detected SMGs are consistent with essentially all redshifts below $z\sim 3$, given a reasonable thermal spectral energy distribution (SED).  The fourth source, the strong radio source SMM-RCS0224.1, has a ratio which appears to place it at $z>3$; however in  this case it is likely the 850$\micron$ flux is also boosted by the AGN, though to a lesser extent than the radio flux, and therefore the 450$\micron$-850$\micron$ flux ratio cannot be described by a simple thermal SED.

To summarize, sources SMM-RCS0224.1, SMM-RCS0224.2 are likely cluster members, based on their optical photometric redshifts, $K$ magnitudes and optical/NIR colors.    Sources SMM-RCS0224.3 and SMM-RCS0224.4 are more consistent with belonging to the background population of SMGs, though  they are not highly lensed by the cluster.  Very little constraint can be placed on the redshift of SMM-RCS0224.5; there are not enough data for an optical photometric redshift estimate and while its radio-submillimeter ratio suggests a $z<$ 1 galaxy, there is substantial uncertainty in this technique, as evidenced by Fig.~\ref{redshift}.   Spectroscopic redshift measurements are clearly required to resolve this issue, as well as improved positional information for the SMGs with large offsets to their radio identifications.  The latter is particularly important to rule out the possibility of single galaxy, or cluster-substructure, lensing.

If two of the SMGs in the  RCS0224 field are indeed dusty starbursts and/or  AGN within the  cluster this has interesting implications  for the evolution of galaxy clusters.  Thus far, star formation rates and dust levels great enough to produce strong enough submillimeter emission to be seen by SCUBA ($\sim$100 M$_\odot$ yr$^{-1}$, but very dependent on temperature) have not been unambigously  detected in clusters.  This could, in part, be a selection effect: in addition to  the fact that very few high redshift galaxy clusters have been searched for star formation, such objects may be so obscured that they go unrecognized in  optical studies, as was the case in the field until recently.  Submillimeter surveys are sensitive to much colder dust than mid-IR observations with ISO, and thus may detect fundamentally different objects. In fact, a deep SCUBA image has been obtained for A1689 but none of the ISO sources are coincident with the 850$\mu$m-selected objects (Knudsen et al.~in preparation). 

We can estimate the star formation rate such systems would have if they lie  at $z=$ 0.773 through an extrapolation from $S_{850\micron}$ to $L_{FIR}$ and using  the relation of  \citet{ken98}.  We adopt a temperature  of 40K  and dust emissivity $\beta$ = 1.3 \citep{dun00}. Ignoring SMM-RCS0224.1, which is likely AGN-dominated, we find a range in star formation rates of $\sim$ 250-430 M$_\odot$ yr$^{-1}$, for our range in 850{\micron} flux levels. We note, however, that this estimate is highly dependent on temperature  \citep[see][for discussion]{eal00}  and varying the dust temperature from 30-50K (a relatively modest range) results in SFR estimates from 90-900M$_\odot$ yr$^{-1}$.   

We note that  both of the clusters in  the literature with substantial 15$\mu$m emission \citep{fad00,duc02} show signs of recent or ongoing cluster mergers, as do many of the clusters with excess radio sources at high redshift and within the local universe \citep{sma99,mil03,gia04}. Currently there is no evidence that RCS0224 is undergoing a merger or large accretion, other than a shallow X-ray map showing  hints of substructure  \citep{hic04},  but this is simply a reflection of the limited amount of data acquired for this cluster so far.  Interestingly, an increase in cluster merging at $z\sim$ 1 also provides an explanation for the super-lensing phenomena, of which RCS0224 is an example. \citet{tor04} show that during a merger the lensing cross-section of a cluster is temporarily enhanced to such a degree that this process can account for the excess multiple arcs seen in optical surveys.   In such a picture, the presence of strong optical arcs may mark merging clusters in which we may also find enhanced star formation and AGN activity.

\section{Conclusions}

We are undertaking an 850{\micron} imaging  survey of z$\sim$1 galaxy clusters selected from the Red-Sequence Cluster Survey.   Here we present the submillimeter data on the first completed cluster, RCS0224, and follow-up imaging for counterpart determination. We detect five SMGs within the field which our mass model shows are not strongly lensed,  and measure  cumulative number counts that  are in general agreement with that expected from other blank-field surveys. Interstingly, this field and two of the three other cluster surveys lie systematically above the blank-field counts, all at roughly 1$\sigma$, and taken together suggest that SMG counts in cluster fields may be enhanced by unrecognized submillimeter luminous cluster members.   A larger sample of clusters will provide tighter statistics.

All five SMGs in the RCS0224 field are detected at  1.4GHz or 450{\micron}, or both,  and we have located NIR counterparts for all the radio detections. Based on accurate optical photometric redshifts, and magnitude and color considerations, we find that two of the five SMGs (SMM-RCS0224.1, SMM-RCS0224.2) are consistent with being cluster member while another two objects are more likely background objects. The redshift of the last source (SMM-RCS0224.5) is poorly constrained; while its radio-submillimeter flux ratio  is consistent with the cluster redshift the large uncertainty in this technique makes this measurement unreliable. Further data, such as spectroscopic redshifts and improved submillimeter positions, will  verify this picture.

\acknowledgements

We thank the anonymous referee for a careful reading of the manuscript.
Research by Tracy Webb is supported by the NOVA Postdoctoral Fellowship program. The JCMT is operated by the Joint Astronomy Center on behalf of the Particle Physics and Astronomy Research Council in the UK, the National Research Council of Canada, and the Netherlands organization for Scientific Research.

\end{document}